\documentclass[aps,prx,twocolumn,superscriptaddress,longbibliography]{revtex4-1}
\pdfoutput=1

\usepackage{graphicx}
\usepackage{color}
\usepackage{amssymb}
\usepackage{multirow}

\usepackage{amsmath}
\usepackage[utf8]{inputenc}
\usepackage{hyperref}
\usepackage{url}
\usepackage{hyphenat}

\newcommand\Fig[1]{Fig.~\ref{#1}}
\newcommand\Tab[1]{Tab.~\ref{#1}}
\newcommand\rmi[2]{\ensuremath{{#1}_{\text{#2}}}}

\newcommand\Vellip{\ensuremath{V_\alpha}}
\newcommand\Vmin{\rmi{V}{min}}
\newcommand\Vmax{\rmi{V}{max}}
\newcommand\Vcell{\ensuremath{V}}

\newcommand\PFloc{\ensuremath{\phi_{l}}}
\newcommand\PFcryst{\ensuremath{\phi_\text{cry}}}
\newcommand\PFlocmax{\ensuremath{\phi_\text{max}}}
\newcommand\PFglob{\ensuremath{\phi_g}}
\newcommand\mean[1]{\ensuremath{\langle {#1}\rangle}}
\newcommand\gparam{m}

\definecolor{orange}{RGB}{255,158,0}
\definecolor{yellow}{RGB}{249,249,0}
\definecolor{blue}{RGB}{0,0,128}
\definecolor{lightblue}{RGB}{44,115,248}
\definecolor{green}{RGB}{0,128,0}
\definecolor{lightgreen}{RGB}{0,255,128}
\definecolor{red}{RGB}{220,20,60}
\definecolor{darkgray}{RGB}{80,80,80}
\definecolor{lightgray}{RGB}{220,220,220}
\definecolor{pink}{RGB}{255,128,255}

\begin{document}

\title{Cuddling Ellipsoids: Densest local structures of uniaxial ellipsoids}

\author{Fabian M. Schaller}
\affiliation{Institut f\"ur Theoretische Physik, FAU Erlangen-N\"urnberg, 91058 Erlangen, Germany}

\author{Robert F.~B.~Weigel}
\affiliation{Institut f\"ur Theoretische Physik, FAU Erlangen-N\"urnberg, 91058 Erlangen, Germany}

\author{Sebastian C. Kapfer}
\email{sebastian.kapfer@fau.de}
\affiliation{Institut f\"ur Theoretische Physik, FAU Erlangen-N\"urnberg, 91058 Erlangen, Germany}


\begin{abstract}

Connecting the collective behavior of disordered systems with local structure on the particle scale is an important challenge, for example in granular and glassy systems.
Compounding complexity, in many scientific and industrial applications, particles are polydisperse, aspherical or even of varying shape.
Here, we investigate a generalization of the classical kissing problem in order to understand the local building blocks of packings of aspherical grains.
We numerically determine the densest local structures of uniaxial ellipsoids by minimizing the Set Voronoi cell volume around a given particle.
Depending on the particle aspect ratio, different local structures are observed and classified by symmetry and Voronoi coordination number.
In extended disordered packings of frictionless particles, knowledge of the densest structures allows to rescale the Voronoi volume distributions onto the single-parameter family of $k$-Gamma distributions.
Moreover, we find that approximate icosahedral clusters are found in random packings, while the optimal local structures for more aspherical particles are not formed.

\end{abstract}

\maketitle

\section{Introduction}

\begin{figure}
\includegraphics{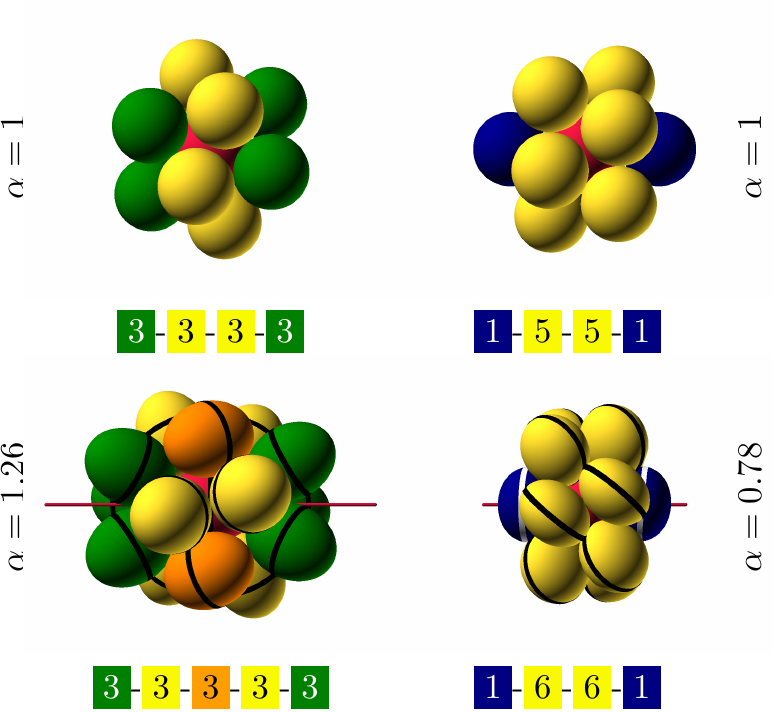}
\caption{ Densest local structures of ellipsoids:
Top row: The 12-neighbor icosahedral cluster of spheres, colored to preserve a 3-fold
rotational axis and three 2-folds (left), and colored to preserve a 5-fold
axis plus five 2-folds (right).
Bottom row: Prolate 15-neighbor and oblate 14-neighbor structures generalizing the above.
Black/white belts mark the equators of the particles.
The red line is the central particle's distinguished axis.
Particles of the same color are related by symmetry operations.
Boxes label the number of particles in the respective rings.
\label{figEyecatcher}
}
\end{figure}

``How many candies can I fit in the jar?''  Children and scientists alike have always been fascinated by packing problems \cite{Weaire2008pursuit}.
Nevertheless, it remains a challenge to understand how collective properties of packings arise from microscopic mechanisms on the particle level.
Microscopic interactions are encoded in local structural motifs comprising a particle and its immediate neighbors.
These structural motifs serve as the building blocks for extended packings. 
A pivotal role in the analytic modeling of granular matter is played by the densest local structure \cite{Aste2008,Aste2007}.
This structure coincides with the solution of the kissing problem in mathematics \cite{Schuette1952,Conway1999}, which asks to maximize the number of spherical particles simultaneously in contact with a central one.
The icosahedral cluster, depicted in \Fig{figEyecatcher}, top row, is the most symmetric
way to arrange the twelve kissers and maximizes the local packing density.
Ideal icosahedra are about 1\% denser than the best \emph{space-tiling} arrangements of congruent spheres,
given by stacked hexagonal lattice planes \cite{Hales2005}.
By embedding distorted icosahedral clusters, disordered granular packings also locally exceed the density limit for space-tiling sphere packings.

\begin{figure*}
\includegraphics{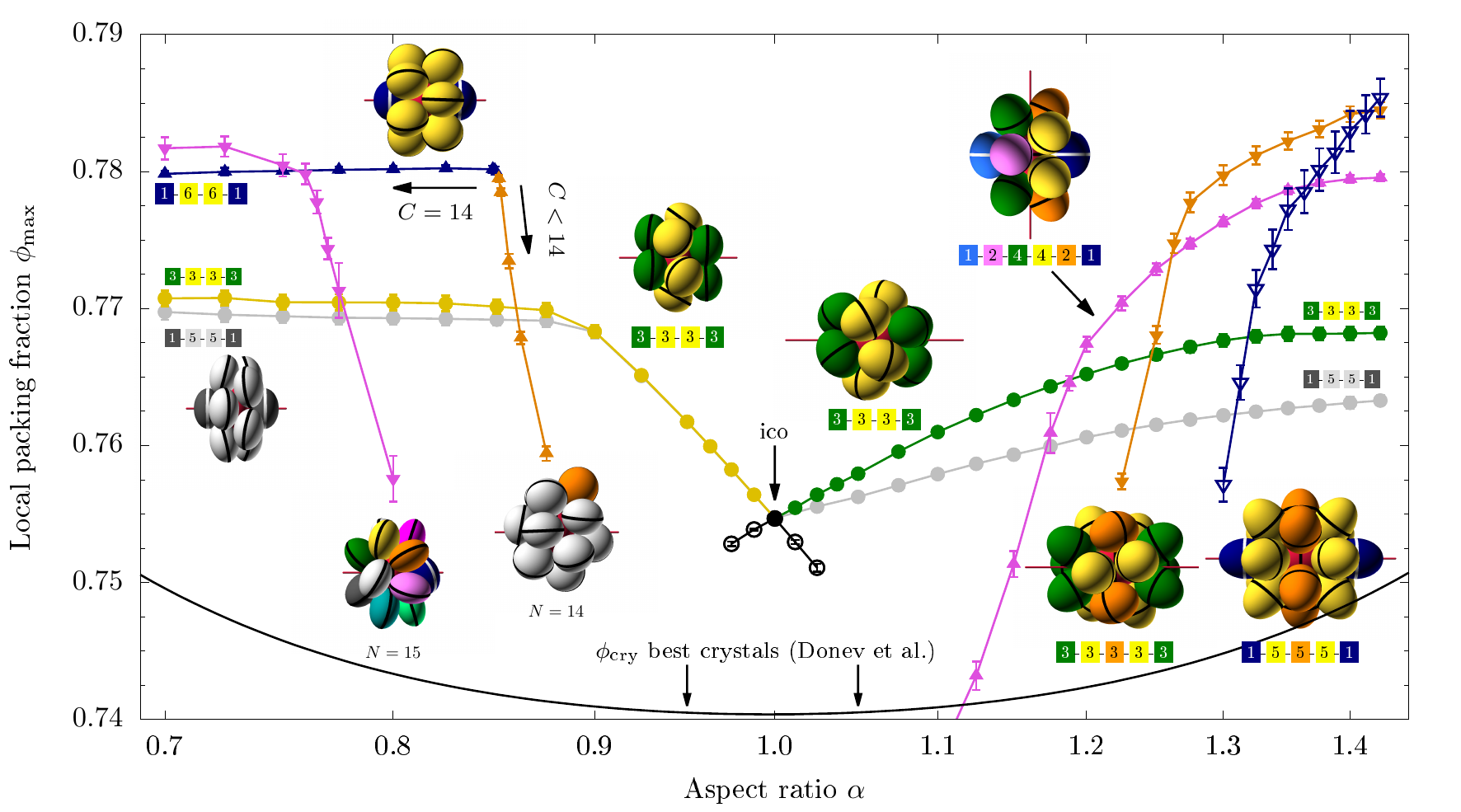}
\caption{Dense local structures of ellipsoids:  Local packing fraction of the central Set Voronoi cell vs.~the
aspect ratio $\alpha$ of the particles.  We identify several branches of candidate structures which are
distinguished by the number and symmetry of Voronoi neighbors.  The plot symbols mark the number of
neighbors ($N=12$: $\bullet$, $N=14$: $\blacktriangle$,
$N=15$: $\blacktriangledown$, $N=17$: $\triangledown$), while different line colors mark a change in symmetry.
The solid line at the bottom of the graph represents the densest-known crystal structure,
and ``ico'' is the icosahedral cluster.
Usually, in the optimal structures, the contact number $C$ is equal to the number of Voronoi neighbors $N$.
One exception is the oblate $N=14$ structure which loses contacts around $\alpha\approx 0.85$ (see kink).
A detailed description of the structures is given in the text.
\label{figMasterplot}}
\end{figure*}

Much like candies, the particles found in nature and in industrial applications are not always spherical and often randomly shaped.
For example, pebbles and sand grains vary widely in size and shape, and
the kissing and packing problems lack rigorous answers.
Generalizations of the sphere-packing problem to congruent aspherical
particles have been intensely studied \cite{DonevCryst2004,Chen2010,Odriozola2013,ChenComplexityPRX2014,Torquato2015lenses}, motivated
both by applications in granular matter \cite{DonevRandom2004,Delaney2010,SchallerPRL2015,SchallerEPL2015,ChaikinTetra2010,Neudecker2012} and by advances in the
synthesis of colloidal particles with prescribed shapes \cite{Glotzer2007anisotropy,Huang2012PolyhedralNanocrystals}.
In general, aspherical shapes pack denser than spheres, and it
could recently be shown that the sphere is a local pessimum for lattice packing
\cite{Kallus2014355,Kallus2}.
A convenient shape for studying the effect of asphericity are ellipsoids.
Even though dense ellipsoid crystals are known \cite{DonevCryst2004},
there is currently no mathematical proof of optimality.

The kissing problem has been generalized to tetrahedra \cite{Lagarias2012MysteriesTetra}, but we are not aware of results for ellipsoidal particles.
Motivated by recent work on the packing properties of disordered granular
ellipsoids \cite{DonevRandom2004,Delaney2010,SchallerPRL2015,SchallerEPL2015}, we here consider a modified
kissing problem for uniaxial ellipsoids $1:1:\alpha$, also known as spheroids, or ellipsoids of rotation.
The number $\alpha$ is the aspect ratio of the particles, that is, $\alpha<1$ ($\alpha>1$) corresponds to oblate (prolate) ellipsoids.
The conventional kissing problem maximizes the number $C$ of particles in contact with a given central particle.
Here, we instead maximize the local packing density $\PFloc = \Vellip / \Vcell$, defined via the volume $\Vcell$ of the central particle's Voronoi cell, where $\Vellip = 4\pi\alpha/3$ is the particle volume. 

The appropriate generalization of the Voronoi diagram for aspherical particles is the Set Voronoi diagram
\footnote{In general, aspherical particles are not fully contained in the Voronoi cells constructed from the particle centers.
For packings of monodisperse spheres, the Set Voronoi diagram coincides with the conventional Voronoi diagram.}, also known as navigational map \cite{Luchnikov1999,Schaller2013}.
The Set Voronoi diagram allocates space by the distance to the particle surfaces instead of the distance to particle centers.
Typical Set Voronoi cells are non-convex and have curved surfaces.
The solution of our modified kissing problem is the structure minimizing the Voronoi volume $\Vcell$.
We define $\Vmin(\alpha)$ as the minimal Voronoi cell volume for each aspect ratio,
and $\PFlocmax(\alpha)=\Vellip/\Vmin(\alpha)$ as the corresponding local packing fraction.
As a first key result of the present article, we numerically determine the optimal local structures for uniaxial ellipsoids
of aspect ratios $0.7\leq\alpha\leq 1.4$.

The description of granular matter and other out-of-equilibrium particle systems by macroscopic variables in the spirit of thermodynamics currently is a nascent field.
A convenient characterization of the particle system is given by Voronoi cell partitions.
In particular, distributions of Voronoi cell volumes are sensitive to structural transitions in granular assemblies
\cite{Finney1970,Starr2002,Lechenault2006,WangSongMakse2010,Schroeder2010epl,Francois2013Geometrical,KlattTorquato2014,SchallerEPL2015}.
The theoretical modeling of such distributions remains an open problem.
For non-interacting particles (ideal gas, Poisson point process), the distribution of Voronoi volumes is almost perfectly fit by a three-parameter Gamma distribution \cite{HindeMiles1980,Okabe2000}.
No fitting curve of comparable accuracy exists for interacting particles, let alone dense packings such as considered here.
Aste et al. \cite{Aste2008,Aste2007} propose a simplified model for sphere packings.
This analytic model yields the full distribution of the Voronoi volumina, but crucially depends on the minimal cell volume $\Vmin(\alpha)$.
Our results for $\Vmin(\alpha)$ permit to extend this model also for random packings of ellipsoidal particles.

In the following section II, we discuss the numerical optimization procedure and the resulting densest local structures.
Section III investigates the occurrence of these motifs as building blocks of extended random packings.
Moreover, we test the analytic model of Aste et al.\ for ellipsoidal particles and discuss future directions.

\begin{figure}
\centering
\includegraphics{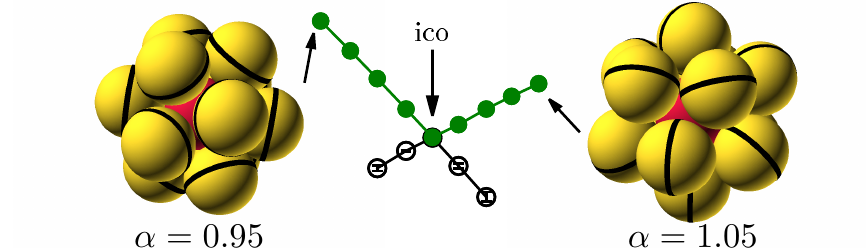}
\caption{Variants of the icosahedral cluster for slightly aspherical ellipsoids.
(left) Neighbors squashed in the normal direction, leading to oblate ellipsoids;
(right) neighbors expanded in the tangential direction, leading to prolate ellipsoids.
The center plot magnifies \Fig{figMasterplot}.
\label{figIcoCrossing}}
\end{figure}

\begin{table}
\centering
\begin{tabular}{l|l||l|l|l|l|l|l|l}
&$\alpha$ & 0.7 & 0.8 & 0.9 & 1.0 & 1.1 & 1.25 & 1.4 \\\hline\hline
\multirow{2}{*}{II.}&$\phi_{\max}$ & 0.781& 0.780& 0.768& 0.755& 0.761& 0.773& 0.784\\
&$N(\phi_{\max})$ & 15 & 14 & 12 & 12 & 12 & 14 & 15 \\\hline
&$\PFglob$ & 0.712 & 0.702 & 0.679 & 0.641 & 0.673 & 0.700 & 0.711 \\
III.&$k$ & 15.4 & 14.9 & 13.3 & 13.5 & 12.3 & 13.8 & 17.4 \\
&$\gparam_1/\gparam_2$ & 0.95 & 0.96 & 0.98 & 0.99 & 0.98 & 0.97 & 0.97 \\
\end{tabular}

\caption{Properties of densest local structures for ellipsoids with aspect ratio $\alpha$ (see section II).
Global packing density $\PFglob$ of disordered Lubachevsky-Stillinger packings analyzed in
section III, and the best-fit parameters of the $k$-Gamma background (see text).
A fraction $\gparam_1/\gparam_2$ of all particles can be attributed to the background.
}
    \label{tabData}
\end{table}

\section{Densest local structures}

Finding the densest local structure of ellipsoids amounts to locating the minimum
in a complex energy landscape, with the energy given by the Voronoi cell volume $V$.
The function $V$ depends on the position and orientation of the $N$ Voronoi
neighbors of the central particle.
In order to solve this optimization problem in $5N$ variables, we use a
two-stage approach:  First, candidate structures are explored by
a simulated annealing scheme; second, the densest candidate structures
found during the annealing phase are optimized by a downhill algorithm.
The numerical optimization considers only a central (red) particle and
the $N$ first-shell neighbor particles that share a Voronoi facet with
it.  Since the optimal number $N$ is not known a priori, we propose
several values of $N$ in separate annealing runs; during one run, the
number $N$ is fixed.  While the red particle is immobile, the other
particles can translate and rotate.  The mobile particles
have a tendency to drift away and detach from the central Voronoi cell.
We thus restrict their
centers to an ellipsoid-shaped arena around the red particle.  Each
trial move proposes to displace and rotate a single mobile particle.
The move is rejected if any overlap between the particles is created.
Otherwise, we compute the volume change $\Delta V$ of the Voronoi
cell belonging to the central particle, and accept the move with
probability $\exp(-\kappa \Delta V)$.  The dimensionless pressure $\kappa$ is $50$ 
initially, and increases to $1000$ over the $2\times 10^6$ Monte Carlo
steps comprising a run.  We keep track of
the best (densest) configuration observed so far, which is restored
at the end of the annealing run.  Finally, we use a downhill algorithm
to find a local optimum of the cell volume.

\begin{figure*}
\centering
\rotatebox{90}{
\includegraphics[width=22cm]{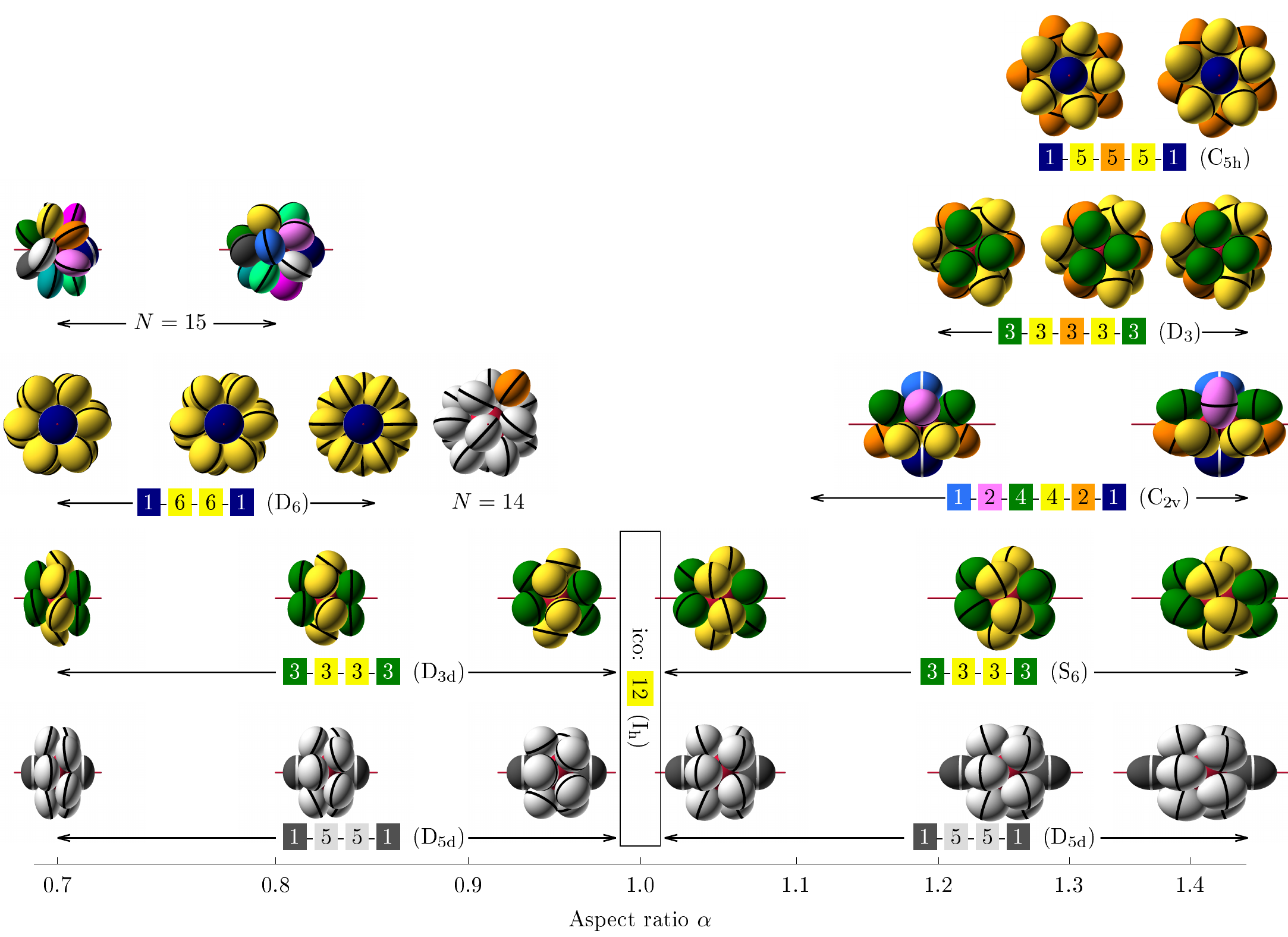}
}
\caption{Evolution of the densest structural motifs with aspect ratio $\alpha$. In addition to the number of particles in each ring (color-coded boxes), we give the point group for each structure. The oblate structures marked $N=15$ and $N=14$ have no nontrivial symmetries.  The \mbox{1-5-5-1} structures (gray) never exceed the \mbox{3-3-3-3}.}
\label{figAppendix}
\end{figure*} 

The key feature of the simulated annealing algorithm is
its ability to escape from local optima, but it does not always return
the global optimum.  We thus perform ten independent
annealing runs for each set of parameters $(N, \alpha)$.  In cases where multiple
structures are in close competition, the number of independent runs was increased to 100.
The major computational expense of our optimization procedure is the
computation of the Set Voronoi cell volume.  To be able to perform sufficiently
many trial moves in the optimization, the resolution of the Set Voronoi
volume computation must be limited.
A typical optimization run uses a discretization \footnote{
For the Set Voronoi computation, we approximate the ellipsoid surface by a set of
seed points for the conventional Voronoi construction \cite{Schaller2013}.
The seeds are first evenly distributed on a sphere, which is scaled to the aspect ratio of
the ellipsoid.  This ensures a higher seed point density where the curvature is large.
} of 126 Voronoi seeds per particle (320 for the downhill phase).
One evaluation of the cell volume takes $\approx 10$~ms (up to 50~ms in the downhill phase)
on a Xeon E5-2630 CPU.  For the final configurations at the end of the downhill phase,
we also compute high-precision cell volumes, and estimate the discretization error
from the difference (see error bars in \Fig{figMasterplot}).

As the result of our numerical optimization scheme, we identify the presumed optimal
structure for each aspect ratio $\alpha$ and Voronoi coordination number $N$, see \Fig{figMasterplot}.
In many of the optimal structures, neighbor particles are related by discrete symmetry operations.
These groups of neighbors share the same color in all figures.  The colored boxes indicate the
number of particles in a ring, with the same color code.
Black and white belts mark the equator of each ellipsoid. The distinguished axis of the
central particle is indicated by the red line.
To obtain more precise coordinates, we also perform constrained
simulations which enforce the rotational and mirror symmetries inferred from some of the
discovered structures (see below).  The reduced degrees of freedom lead to improved
convergence of the simulated annealing algorithm \footnote{
Breaking any of the symmetries does not necessarily increase the
Voronoi cell volume by a large amount, and the unconstrained annealing optimization
does not necessarily find the most symmetric structure.  We have, however,
checked using the Nelder-Mead algorithm that our optimal structures are local minima within
numerical precision.}.

As the particles become more
aspherical, a general trend towards denser structures and increased $N$ exists
(see \Tab{tabData}, row II).
In this sense, the sphere is not only a worst case of the lattice packing
problem \cite{Kallus2014355,Kallus2} but also of the local packing problem.
For practical reasons,
we limit our study to moderately aspherical ellipsoids.  First, the pointed
tips of very aspherical particles necessitate a fine discretization in the Set Voronoi
computation.  Second, dense structures of very aspherical objects strongly depend
on minute details of the particle shape \cite{ChenComplexityPRX2014} and are less
generic than the results for moderate asphericity considered here.

In the densest local structure of spheres \cite{HalesMcLaughlin}, the symmetric icosahedral cluster
(ico; see \Fig{figEyecatcher}, top row), the first-shell neighbors do not
touch each other, leaving space to be filled.  In fact, the twelve kissing spheres
leave so much space that it is possible for them to switch places without
ever losing contact with the central sphere.  Thus, packing density can be increased in two
ways: One can either expand the neighbors in a direction parallel to the central
sphere's surface, leading to prolate ellipsoids, or squash the neighbors in the
normal direction, producing oblates.  Thus, for small asphericity, oblate and
prolate ellipsoids prefer different orientations of neighbors with respect to
the central particle's surface, and pack densest in two different structures,
see \Fig{figIcoCrossing}.  The two branches cross in the ico configuration at $\alpha=1$.

Moving away from the spherical case, the icosahedral symmetry is broken.
Even though the neighbor particles twist, a threefold rotation axis is preserved both for oblates and prolates,
and the densest structure consists of a succession of four three-rings (\mbox{3-3-3-3} structures).
While both oblates and prolates show three twofold axes, the oblates have dihedral mirror planes (point group D$_\text{3d}$),
and prolates preserve an improper sixfold axis (S$_6$).
The evolution of these structures with $\alpha$ is illustrated in \Fig{figAppendix}.
The icosahedral cluster also features fivefold axes, which would imply a \mbox{1-5-5-1}
structure, see \Fig{figEyecatcher} (top right).
The \mbox{1-5-5-1} structure is degenerate with the \mbox{3-3-3-3} for sphere-like oblates, see gray and yellow curves
in \Fig{figMasterplot}.
For all other aspect ratios, the fivefold symmetry is not stable with respect to unconstrained global optimization.

In the bottom row of \Fig{figEyecatcher}, we illustrate two ways to generalize
the ico cluster towards higher $N$:  to introduce additional rings of particles
(left), or to increase the number of particles per ring (right).  Most of the dense
structures found in this study can be categorized in this fashion, see \Fig{figMasterplot}.  For example,
prolate ellipsoids with $1.26 \lesssim \alpha \lesssim 1.42$ pack densest in a structure
with threefold symmetry, composed of five three-rings of particles (\mbox{3-3-3-3-3}, point group D$_3$). At
higher $\alpha$, a structure with fivefold symmetry and $N=17$ is preferred (\mbox{1-5-5-5-1}, C$_\text{5h}$).
For some of the structures, the rotational symmetries of the particle rings are broken.
Consider for example the prolate \mbox{1-2-4-4-2-1} structure with $N=14$ neighbors,
which is optimal in the range $1.19 \lesssim\alpha\lesssim 1.26$.
In this structure, the rings are not stacked along the central particle's axis.
Consequently, the rotational symmetry is lifted by the central particle, and only two mirror planes are preserved (C$_\text{2v}$).

Oblate particles below $\alpha\approx 0.86$ form a 14-coordinated structure instead of the
icosahedral cluster, see \Fig{figMasterplot}.
The full sixfold symmetry is reached only for $\alpha\lesssim 0.85$
(\mbox{1-6-6-1}, D$_6$) where all neighbors touch the central particle (contact number $C=14$).
Above $\alpha\approx 0.85$, due to steric constraints, not all of the Voronoi neighbors can touch the central particle,
hence $C<14$.
The sixfold symmetry is lost and the packing fraction of the $N=14$ structures declines rapidly.
Once the packing fraction falls below the $N=12$ branch, the cluster tends to expell a particle (orange particle in \Fig{figMasterplot}).
For very oblate ellipsoids, $\alpha\lesssim 0.76$, fifteen neighbors without any obvious symmetry pack densest around the central particle (labeled $N=15$ in \Fig{figMasterplot} and \Fig{figAppendix}).
We conjecture that for even larger asphericities than considered here, the densest structures are typically disordered without any symmetries.

\section{Dense random packings}

Having established the densest local structures for each aspect ratio, we proceed to analyze the occurrence of these motifs in dense disordered packings.
We generate, for each aspect ratio, at least 100 random packings of 5000
monodisperse ellipsoids each using the Lubachevsky-Stillinger-Donev protocol \cite{Donev-LSD-1,Donev-LSD-2}.
The packing generation protocol does not implement gravity, and consequently
the packings are essentially free of orientational order \footnote{
The nematic order parameter tensor is close to isotropic, its largest eigenvalue is smaller than 0.04.}.
We set the expansion rate of the particles to $3\times 10^{-6}\cdots 10^{-5}$ times the thermal velocity,
which produces dense packings, but avoids the eventual formation of crystalline domains.
Thus, our packings are representative of the random close-packed (or maximally random jammed) state for the respective aspect ratio \cite{DonevRandom2004,Delaney2010}.
\begin{figure}
\includegraphics{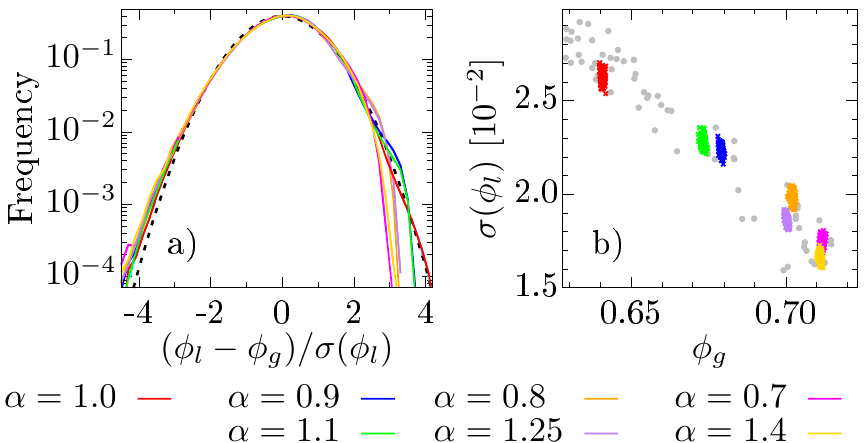}
\caption{
a) Distributions of the rescaled variable $(\PFloc-\PFglob)/\sigma(\PFloc)$.
For reference, a Gaussian distribution is shown (black dotted curve).
b) Behavior of the standard deviation of local packing fractions $\sigma(\PFloc)$.
The gray points are reference data from Ref.~\cite{SchallerEPL2015}.
}
\label{figiPfDistr}
\end{figure}
Our data is consistent with published tomography experiments and numerical sedimentation data for oblate ellipsoids \cite{SchallerEPL2015}.
\Fig{figiPfDistr}a shows the distributions of local packing fraction.
Due to high statistics, we can resolve the fringes of the distribution which are not yet accessible in experiment.
The standard deviations $\sigma(\PFloc)$ are also in very good agreement with earlier data for oblate ellipsoids (gray data points in \Fig{figiPfDistr}b).
Moreover, \Fig{figiPfDistr}b demonstrates a surprising symmetry between ellipsoids of reciprocal aspect ratios:
Both global packing fractions $\PFglob = 1/\mean{1/\PFloc}$ and the standard deviations $\sigma(\PFloc)$ approximately coincide.
This finding continues a series of unexplained correspondences between oblate and prolate ellipsoidal particles, for example the equilibrium phase diagram \cite{Odriozola2013} and packing fractions \PFcryst\ of the best known crystals \cite{DonevCryst2004}.

\begin{figure}
\includegraphics{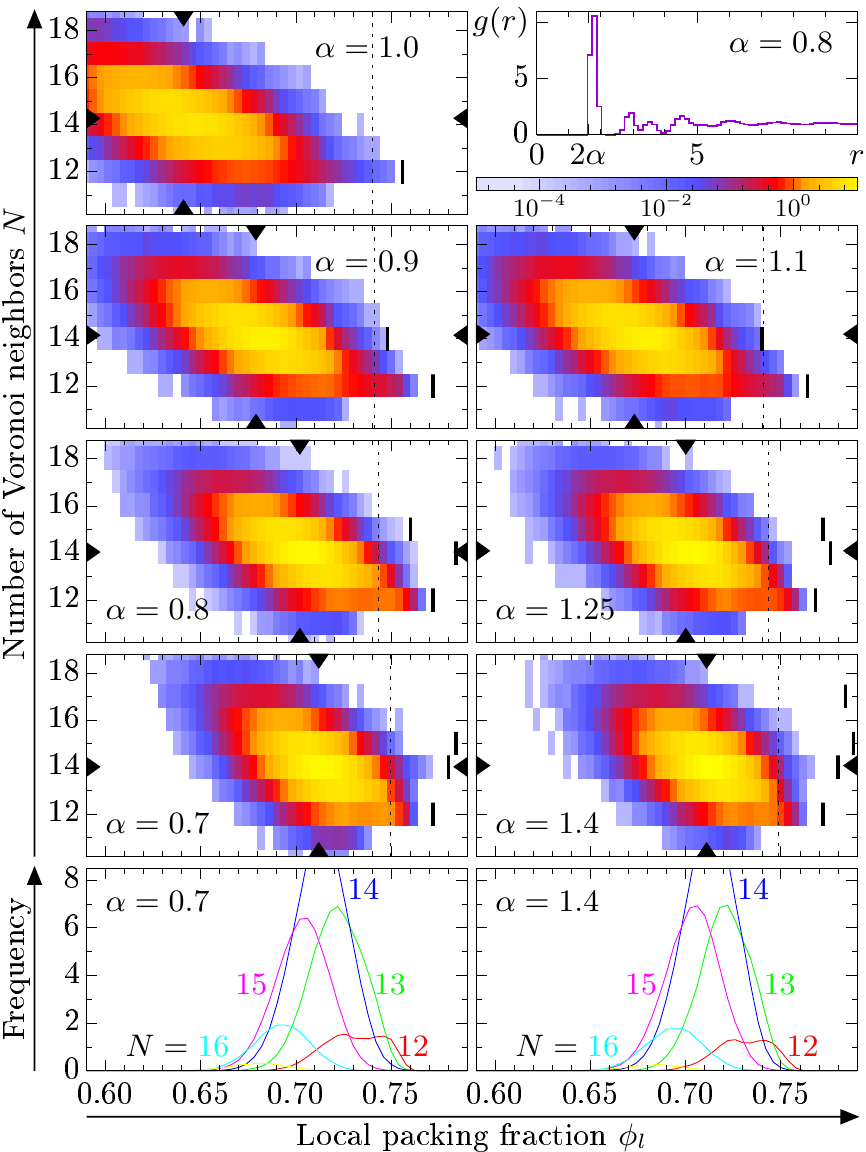}
\caption{   Distribution of Voronoi neighbor count and local packing fraction.
            The bold vertical bars indicate the densest local structures for each $N$ (see \Fig{figMasterplot}.
            The vertical dashed line marks the densest known crystalline packing $\PFcryst$ \cite{DonevCryst2004},
            and the triangular tics on the axes mark the global packing fraction and mean Voronoi neighbor count
            of the dense disordered packings.
            Bottom row: Marginal distributions of the above for different $N$.
            Top right plot: Radial distribution of the supracrystalline clusters, $\PFloc > \PFcryst$.
}
\label{figNNcount}
\end{figure}

\Fig{figNNcount} displays the frequency of Voronoi cells in our dense disordered
packings with a given coordination number $N$ and packing fraction $\PFloc$.
The bold vertical bars indicate, for each Voronoi coordination number, the maximal packing fraction found in \Fig{figMasterplot}.
At small asphericity, the densest local structure is the icosahedral cluster.
Random packings of spheres are known to contain distorted variations of icosahedral clusters \cite{Anikeenko2007}, while the probability for perfect ico clusters vanishes \cite{Schroeder2010epl}.
Our data for $\alpha=0.9, 1.0, 1.1$ in \Fig{figNNcount} confirms these results for packings of moderate asphericity.
More aspherical ellipsoids ($\alpha\leq 0.8$ and $\geq 1.25$) could pack denser with fourteen or more neighbors.
As \Fig{figNNcount} demonstrates, such structures are not formed in significant amounts.
Instead, the densest clusters are again $N=12$ cells.
Visual inspection and a quantitative analysis using Minkowski structure metrics \cite{Mickel2013} (see appendix) confirm that
very aspherical particles tend not to form their optimal structures, but do form distorted variations of the $N=12$ optimal structures.
The densest clusters in a random packing readily exceed the best known ellipsoid crystals of packing fraction $\PFcryst$.
These supracrystalline clusters ($\PFloc>\PFcryst$) are almost uniformly distributed in the packing.
The presence of a supracrystalline structure implies reduced packing fraction in its vicinity as these motifs cannot be periodically continued.
The radial distribution function $g(r)$ of supracrystalline clusters is reminiscent of a hard-core fluid with short-range repulsion and quickly decays to unity, see top right plot in \Fig{figNNcount}.
We return to this point in the conclusion.

While Voronoi diagrams are a powerful tool for characterizing granular packings, there is currently no established theory for aspherical particles.
Aste et al.\ propose an analytic model \cite{Aste2008,Aste2007} which predicts,
under simplifying assumptions, the full distribution of Voronoi cell volumina.
They find a so-called $k$-Gamma distribution, with probability density
\mbox{$\Gamma_k (x) \propto x^{k-1} \textnormal{exp}(-kx)$, }
where the quantity $x$ is the re-scaled volume 
\begin{align}
    x = \frac{V-\Vmin(\alpha)}{\langle V\rangle-\Vmin(\alpha)}\;.
\label{eqAste}
\end{align}
For jammed spheres, the parameter $k$ varies between 11 and 15, which agrees with typical neighbor counts \cite{Aste2008}.
The value of $k$ has been linked to a granular temperature \cite{Aste2007}.
The shape of the $k$-Gamma distributions strongly depends on $\Vmin(\alpha)$, which was previously unknown for ellipsoids.

\begin{figure}
\includegraphics{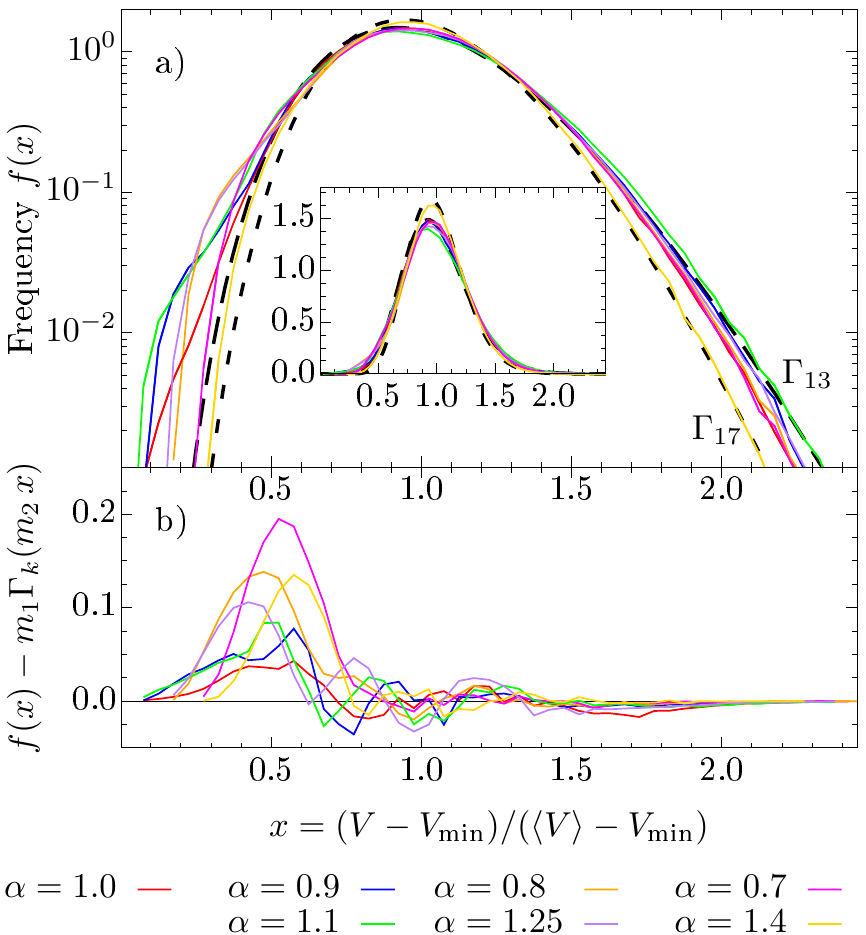}
\caption{%
Distribution of rescaled Set Voronoi cell volumina in random ellipsoid packings.
a) Data for different aspect ratios $\alpha$, compared with two $k$-Gamma distributions with $k=13$ and $k=17$.
The inset shows the same data on a linear scale.
b) Residual after subtracting a best-fit $k$-Gamma distribution (see text).
}
\label{figDistr}
\end{figure}

Combining our present results for $\Vmin(\alpha)$ and  the $k$-Gamma model,
we can now predict the full distribution of Voronoi volumina, with only a single parameter $k$.
\Fig{figDistr}a contrasts our data for dense random packings with the predictions from the $k$-Gamma model.
Both for spheres and ellipsoids, the model curves satisfactorily reproduce the
skewed distributions of the data.
Upon closer inspection, however, the data exhibit a shoulder at small volumes which deviates from the $k$-Gamma curve.
The position of the shoulder is lowest for spheres, and shifts into the main peak (larger $x$) for aspherical particles.
A similar feature is also present in packings of bidisperse disks (see Fig.~2 and 3 in Ref.~\cite{Lechenault2006})
and random-close-packed colloidal particles (see Fig.~4 in Ref.~\cite{KuritaWeeks2010}).

In order to isolate the origin of the shoulder contribution, we fit an unnormalized $k$-Gamma probability density,
$\gparam_1 \Gamma_k (\gparam_2\, x)$, to the upper portion of the data ($x>0.6$ or $x>0.8$, depending on $\alpha$).
The majority of Voronoi cells, $\gparam_1/\gparam_2\geq{}$95\%, can be attributed to the $k$-Gamma-distributed background.
For our packings, the best-fit value for $k$ is between 12 and 17, see \Tab{tabData}, row III.
In general, $k$ increases with asphericity, while the average Voronoi neighbor count stays almost constant, $\mean{N}\approx 14$.
\Fig{figDistr}b shows the residual after subtracting the $k$-Gamma background from the Voronoi volume distributions.
We find an excess of Voronoi cells at low volumes, indicating that there are `attractor' motifs at these packing fractions which are preferentially formed.
This excess mainly stems from $N=12$ cells which show a bimodal $\PFloc$ distribution (see \Fig{figNNcount}, bottom row).
Evidently, the attractors which cause our packing to deviate from the $k$-Gamma model are the $N=12$ motifs for the relevant aspect ratio.
Excluding such specific packing motifs, the $k$-Gamma model provides useful and accurate parametrization of dense frictionless ellipsoid packings.

\section{Conclusion}

In the present article, we explored the influence of asphericity $\alpha$ on local packing motifs.
Opening a new chapter in the story of the classical kissing problem, we propose a modification for arbitrarily-shaped particles which is tractable by numerics. {We present the results of an extensive numerical search for optimal structures.}
Surprisingly, many of our densest structures exhibit a high degree of symmetry.
With increasing asphericity, we find a trend towards higher coordination numbers and packing fractions.
Even though we do not maximize the contact number, in most of our optimal structures,
all neighbor particles touch the central one, i.\,e., $C=N$.
One exception is the oblate $N=14$ structure which loses contacts for particles too spherical (see kink in \Fig{figMasterplot}).
Our optimal structures also are good candidates for solving the classical kissing problem for ellipsoidal particles.
A numerical approach similar to ours was recently employed to study clusters of Platonic solids with interesting physical properties \cite{Teich09022016}.
The continuous degree of freedom present in our ellipsoid clusters could be useful in realizing new kinds of symmetries, and tailoring the structure to applications.

Furthermore, we connect these densest motifs to the properties of disordered ellipsoid packings.
Knowledge of the minimal cell volume $\Vmin(\alpha)$ permitted us to rescale our data of aspherical particle packings onto the curves predicted by the $k$-Gamma model.
The rescaled distributions of Voronoi volumina are well described by the model curves.
The remaining small deviations can be explained by the excess formation of certain types of $N=12$ clusters.
The success of the $k$-Gamma model is unexpected as it includes correlations between neighboring particles only via the minimum packing volume.
In particular, mechanical stability of a packing would also imply an \emph{upper} cutoff $\Vmax$ for the Voronoi cell volume distribution, not included in the model.
$\Vmax$ would also depend on additional microscopic parameters such as friction and hence would be even more difficult to establish than $\Vmin$.
We expect that $\Vmax$ will be essential for the description of loose packings;  in the present random-close-packed limit, low compactivity effectively masks this effect.

The local packing motifs studied here can be regarded as the building blocks of disordered packings, and are analogous to the atoms in a solid.
The interaction between neighboring `atoms' is of similar complexity as the energy landscape of glasses,
and the treatment of disordered phases in granulates, glasses and complex fluids remains a challenging task.
However, the frequency and spatial distribution of supracrystalline motifs may provide some insight into the hidden structure of packings.
Interestingly, only variations of icosahedral clusters are preferentially formed in our random packings.
The optimal structures of our more extreme ellipsoids are not realized in random packings, an effect which is increasing with asphericity.
One could imagine that slight polydispersity might amplify the formation of optimal motifs.
Supracrystalline clusters exhibit nontrivial correlations on length scales above the particle scale (see inset in \Fig{figNNcount}) which require further investigation.
It will be interesting to see whether and how the history of packing formation, such as sedimentation under gravity or compaction by shaking are reflected in these signatures, and what additional information they can reveal about the architecture of the packing.

As demonstrated here, ellipsoids are a useful testing ground for the effects of variations in grain shape.
Our findings are immediately relevant for the realistic modeling of granular materials consisting of aspherical, polydisperse and randomly-shaped grains.
Such materials are found in geological processes such as the dynamics of dunes or tali, and industrial applications.

\section*{Acknowledgments}

We acknowledge funding by the German Science Foundation (DFG) through the research group `Geometry and Physics of Spatial Random Systems' under grant SCHR-1148/3-2, and by the cluster of excellence `Engineering of Advanced Materials'.
We thank Aleksandar Donev for support with his Lubachevsky-Stillinger code, and Markus Spanner for providing a large dataset of jammed spheres.
Finally, we are grateful to the anonymous referees, Gerd Schröder-Turk, Matthieu Marechal, Johannes Hielscher, and Klaus Mecke for advice and useful comments.

\bibliography{lit}
\vspace*{1cm}

\appendix

\section*{Appendix: Morphometric structure analysis of random packings}
\begin{figure*}
\includegraphics{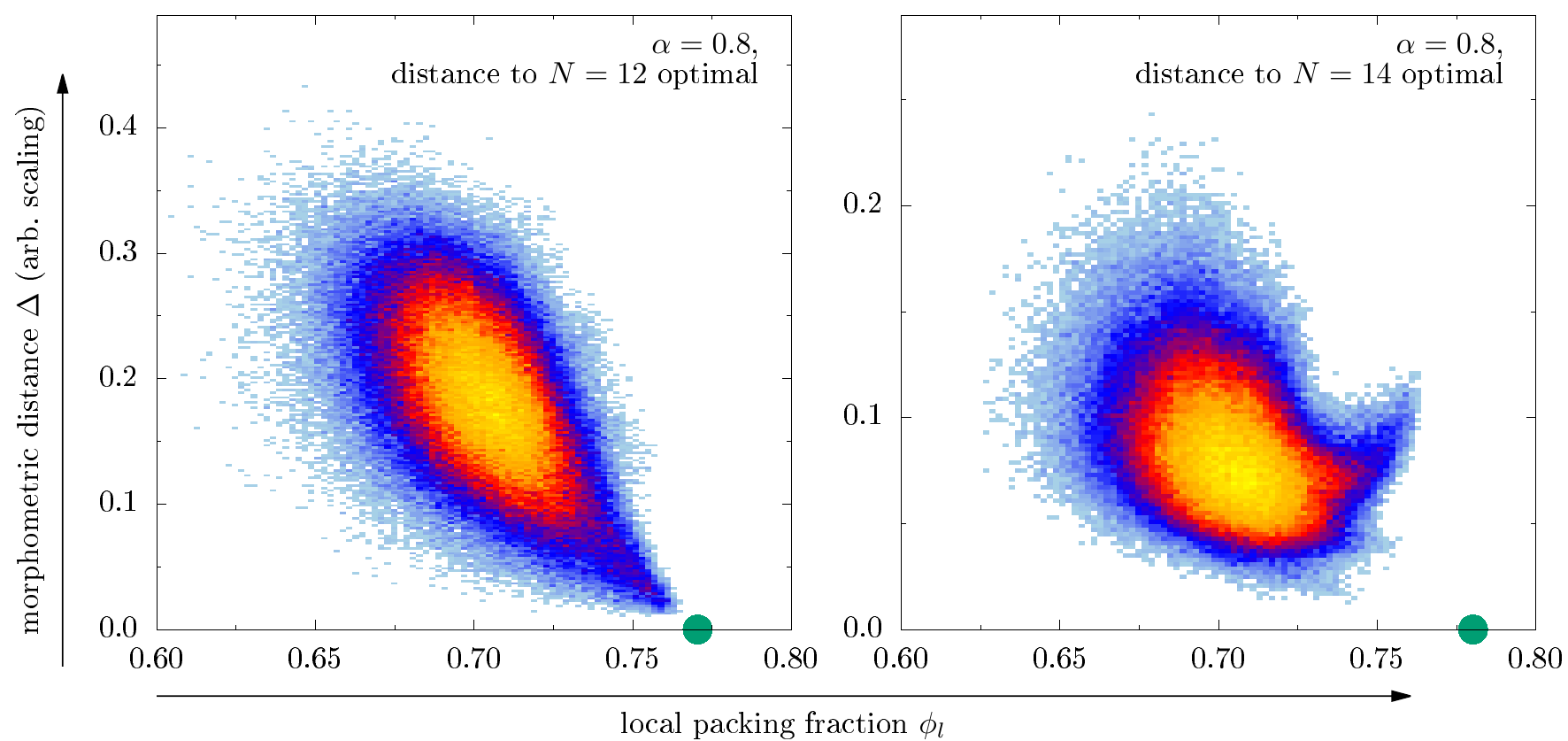}
\caption{
   Morphometric analysis of the dense random packings of $\alpha=0.8$ ellipsoids.
The horizontal axis shows the local packing fraction of the individual Set Voronoi cells.
The vertical axis shows the morphometric distance $\Delta$ of each cell from the cell of two of our optimal (densest) structures.
Left:  Clusters in the packing approach the \mbox{3-3-3-3} structure (green bullet).
Right:  There is no trend towards the \mbox{1-6-6-1} structure (green bullet). 
    }
    \label{figMorphoplot}
\end{figure*}

The appearance of distorted variation of the $N=12$ optimal structures can be demonstrated
by a morphometric analysis using Minkowski structure metrics \cite{Mickel2013}.
We compute, for each Set Voronoi cell $K$ in the random packing,
the Minkowski structure metrics (MSM) $q_l$, defined via
\begin{align}
    q_{lm}(K) &:= \frac 1{A(K)} \int_{\partial K} \! \mathrm d^2 r \; \; Y_{lm}(\theta, \varphi)
\\
    q_l(K) &:= \sqrt{\frac{4\pi}{2l+1} \sum_{m=-l}^{l} | q_{lm} |^2}
\end{align}
where $A(K)$ is the surface area of the cell $K$, $\partial K$ is its boundary,
and $Y_{lm}$ are spherical harmonics with
the spherical coordinates $\theta, \varphi$ taken with respect to the particle center of mass.
The MSM are translation- and rotation-invariant metrics of the shape of the individual Voronoi cells.
The distance of each individual Set Voronoi cell from a reference structure
can be quantified by considering the pseudo-distance function
\begin{align}
    \Delta(K) := \sum_{l=2}^7 \bigl(q_l(K) - q_l(\text{ref})\bigr)^2
\end{align}
with the MSM for the reference structure, $q_l(\text{ref})$.
\Fig{figMorphoplot} shows the distribution of the morphometric distances $\Delta$ and local
packing fractions \PFloc\ in random packings of $\alpha=0.8$ ellipsoids (other aspect ratios
show the same result).
On the left, the reference structure is the \mbox{3-3-3-3} structure for $\alpha=0.8$.
A continuum of local structures approaches the reference motif (green bullet on the horizontal axis).
On the right-hand side, the reference structure is the optimal \mbox{1-6-6-1} motif.
The particles show no propensity to form this structure in a random packing.

\end{document}